\documentclass[12pt,a4paper]{article}
\usepackage{float}
\usepackage{amsmath}
\usepackage{amstext}
\usepackage[psamsfonts]{amssymb}
\usepackage[unicode,bookmarks,bookmarksopen,bookmarksopenlevel=0,colorlinks,%
plainpages=false,pdfpagelabels,hypertexnames=false,pageanchor=true,breaklinks=true,%
linkcolor=blue,citecolor=blue,urlcolor=red]{hyperref}
\usepackage[left=2.0cm, right=2.0cm, top=1.0cm, bottom=1.5cm,
includehead]{geometry}
\usepackage{epsfig}
\usepackage{upref}  %ссылки независимо от контекста печатаются прямыми %
\usepackage{indentfirst}
\usepackage[mathcal]{euscript}
\usepackage[cp1251]{inputenc}
\usepackage{version}

\newcommand\ben{\begin{enumerate}}
\newcommand\een{\end{enumerate}}
\newcommand\bed{\begin{itemize}}
\newcommand\eed{\end{itemize}}
\newcommand\bei{\begin{description}}
\newcommand\eei{\end{description}}
\newcommand\be{\begin{equation}}
\newcommand\ee{\end{equation}}
\newcommand\beq{\begin{eqnarray}}
\newcommand\eeq{\end{eqnarray}}
\newcommand\beqo{\begin{eqnarray*}}
\newcommand\eeqo{\end{eqnarray*}}

\newtheorem{Exmpl}{Example}

\newcommand\const{\operatorname{const}}

\def\eps{{\epsilon}}

\def\be{{\bf e}}

\newcommand\pt{\partial}

\newcommand\RE{\operatorname{Re}\,}

\begin{document}
%\textbf{\emph{May, 23. 2018.}}
\thispagestyle{empty}
\begin{center}
{\large \textbf{A REMARK ON THE DISORIENTING   OF SPECIES DUE TO}\smallskip\\
{\bf FLUCTUATING ENVIRONMENT}}

\bigskip
{\large \textbf{Andrey Morgulis}\footnote{corresponding author}}
\\
{Southern Mathematical Institute of VSC RAS, Vladikavkaz, Russia\\
 I.I.Vorovich Institute for Mathematic, Mechanics and Computer Science,\\
  Southern Federal University, Rostov-na-Donu, Russia}
\\
\texttt{{Email:morgulisandrey@gmail.com}}

\smallskip
{\large and}

\smallskip
{\large \textbf{Konstantin Ilin}}
\\
{Dept. of Math, The University of York, Heslington, York, UK}
\\
{\texttt{Email:konstantin.ilin@york.ac.uk }}

\bigskip
\begin{minipage}{150mm}
\noindent
{\footnotesize
\begin{center}
\textbf{Abstract.}
\end{center}
In this article, we study the short-wavelenght stabilizing of  a cross-diffusion system of Patlak-Keller-Segel (PKS) type. It is well-known that such systems are capable of behaving   rather  complex way due to the destabilization and bifurcations of  more simple  regimes. However, such transitions (as far as we aware) have been studied only for homogeneous equilibria of homogeneous (i.e. translationally invariant) PKS systems. In the present article, we get rid of the  translational invariance by assuming that  the  system is capable of processing an external signal.
%consider a system that does not possess the  translational invariance because of the reacting to %For instance, the external signal may  arise from the inhomogeneity of the distribution of an environmental characteristic such as   temperature, salinity, terrain relief, etc.
%In the case of homogeneous distribution  the model restores the translational invariance  since such distribution produces  no  signal.
%Our goal is to examine the effect of  the short-wavelength signals. To do this, we employ a homogenization.
%We separate the short-wavelength and smooth components of the system's response and derive a ‘slow' system governing the latter one.  %The slow system takes almost the same form as the original system  but  the  inhomogeneity makes its own contribution  into the flux of the predators. This additional flux vanishes  when a characteristic amplitude of the inhomogeneity  is equal to zero; that is, the slow system includes the case of homogeneous environment as a particular case.
We examine the effect of short-wavelength signal with the use of homogenization. The  homogenized system evinces an exponential reduction of cross-diffusive transport in response to the  increase in the external signal  level.  Such the loss of cross-diffusive motility, in turn, stabilizes the primitive quasi-equilibria and  prevents the occurrence of  more complex unsteady patterns to a great extent.}
\end{minipage}
\end{center}
\section*{Introduction}
\label{Intro}\noindent
Taxis is usually defined as an ability of a biological substance to respond to another substance, called stimulus or signal, by directional motion on a macroscopic scale. In particular,  so-called chemotaxis is  driven by chemical signals. The well-known Patlak-Keller-Segel (PKS) model assumes that the chemotactic flux of species is directed along the gradient of stimulus and in this sense represents a non-linear cross diffusion. The PKS approach is widely used for the modelling of the other forms of taxis. For example, the stimulus for one species (the ‘predators’) may be the density of another species (the ‘prey’) or some other signal emitted by the ‘prey’. This may be some kind of chemical or something else which is either attractive or repellent for the ‘predators’. Such interaction of species is known as prey-taxis \cite{Ts94}-\cite{Ts04-1}, \cite{TllWrzk,TtnZgr,LiTao}. For more insights into the PKS systems and their applications,  one can refer to  articles \cite{GMT,AGMTS,Chaplain,BellBellTao,QiJiLu,Black1}  and also follow  the references given in there.

%One can found Various  insights into  the PKS systems and their applications, and   references on the subject  can be found in
%In the present study we consider so-called indirect prey-taxis when the stimulus for the `predators' is not the `prey'  density itself but a signal emitted by the `prey'.

%The present study concerns  so-called prey-taxis or chemotaxis. In the former case,; in the latter case  the  density of the `prey' is replaced by a signal produced by the `prey'. This may be a chemical  released by the `prey' or something else  which can be  either attractive or repellent for the `predators'. The second kind of the species interaction can be also classified indirect prey-taxis.

It is well-known that PKS  systems are capable of behaving   rather  complex way due to the destabilization and local bifurcations of  more simple  regimes. However, such transitions (as far as we aware) have been studied only for  for homogeneous equilibria\footnote{ In such an equilibrium, the distributions of all species are supposed to be homogeneous} of homogeneous (i.e. translationally invariant)  systems \cite{GMT,AGMTS,Chaplain,TtnZgr,QiJiLu,LiWangShao}.  The effect of spatial-temporal inhomogeneity is still rarely addressed in the literature.  We know only of  articles \cite{YrkCbbld} and  \cite{Black1,IssShn}. The former  aimed to the modelling of the effect of the terrain relief on the spatially distributed living community studies a homogenization of a reaction-diffusion system with  jumps of  coefficients across points of a mesh when this mesh getting finer. The latter two articles treat the issues of global boundedness of solutions.

In the present article, we consider a system without translational invariance. This system   is resulted from the adding of an external signal to the model originally proposed in \cite{GMT,AGMTS} for a predator-prey community with prey-taxis. More precisely, we assume that, in addition to the prey-taxis, the predators are endowed with taxis  driven by the external signal.  As the signal, they can perceive, for example,   an  inhomogeneity in the distribution of some environmental characteristic such as  temperature, salinity,  terrain relief, etc.

%breaking  of the translational invariance due to an external signal which is applied to the model originally proposed in \cite{GMT,AGMTS} for a predator-prey community with prey-taxis. It is perhaps the most simple PKS-type system  capable of transiting from the homogeneous  equilibria to self-oscillatory wave motions.
%by assuming that the system is capable of handling an external signal.
%system arising from the adding of an external signal into the model originally proposed in \cite{GMT,AGMTS} for a predator-prey community with prey-taxis.  with no translational invariance.

%of the  latter is due to an external signal  which is accepted by the predators in addition to the signal emitted  by the prey and they both are capable of bringing  the predators into moving.
%The external signal may arise, for instance, from an inhomogeneity in the distribution of an environmental characteristic such as  temperature, salinity,  terrain relief, etc. We introduce such external signal  It is perhaps the most simple PKS-type system  capable of transiting from the homogeneous  equilibria to self-oscillatory wave motions.%Our goal is to examine the effect of a short-wavelength signal. For this purpose, we employ a homogenization.

The system introduced in \cite{GMT,AGMTS} is perhaps the most simple PKS-type system being capable of transiting from the homogeneous  equilibria to self-oscillatory wave motions via the local bifurcation.  We stress here that this transition does not involve the predators kinetics but   the taxis only. This is why this system seems to be most suitable  for a primary study of the effects of inhomogeneity on such transitions in the PKS-systems. Here we  focus ourselves upon the short-wavelength signals which we  examine using the homogenization technique, e.g. \cite{Allr-1}-\cite{Allr-3}.  It turns out, that the  homogenized system and homogeneous system, in  which the external signal is off,  are  very similar but  the external signal  induces    additional  drift of predators  which  vanishes when the signal is off. This drift causes a reduction of the predator motility.  Namely, the effective motility  goes down exponentially in comparison with the homogeneous system in response to the increase in  the short-wavelength  signal level. The loss of motility prevents, to a great extent, the occurrence of the waves and dramatically stabilizes the primitive quasi-equilibria fully imposed by the external signal. This fact allows an interpretation: while  processing  the  intensive small-scale fluctuations of the environment the predators  are  unable to pursue the prey effectively, and this can be seen as their disorientation.

The article is organized into five sections supplemented with two Appendices. In section 1, we discuss the governing equations and their relation to the PKS-systems. In section 2, we consider the case of homogeneity and present the results of the stability analysis of
the homogeneous equilibria of homogeneous system which are necessary for comparison with what we get upon the homogenization. In section 3, we pass to the case when the shortwave external signal is on  and we describe the homogenized system. In section 4, we examine the stabilization of quasi-equilibria. Section 5 contains the discussion of the results. Appendix I describes a routine part of the stability analysis. Appendix II contains the details of the homogenization procedure.
\section{The governing equations}
\label{SecInrtTxs}
\setcounter{equation}{0}
Let us consider a dimensionless  system
\begin{eqnarray}
&& \pt_tu=\pt_x(\kappa q+ f) -\nu u+\delta_u\pt^2_x u;
\label{TxEq}\\
&& \pt_tp=\pt_x(\delta_p \pt_xp-pu);
\label{PrdEq}\\
&& \pt_tq=q(1-q-p)+\delta_q \pt^2_x q.
\label{PryEq}
\end{eqnarray}
Here $x,t$ stand for a spatial coordinate and time;  $p$ stands for the predators distribution density; $q$ stands for the prey distribution density; $u$ stands for the macroscopic velocity of the predators; $\partial_t$ and $\partial_x$ denote partial differentiation with respect to $t$ and $x$; $\delta_p,\delta_q,\delta_u$, $\kappa$,  $\nu$ are positive parameters.

Equations (\ref{PryEq}) and (\ref{PrdEq}) describe balances of the prey and of the predators respectively. We assume that  reproduction of the prey  and its losses due to predation  obey the logistic and  Lotka-Volterra laws correspondingly. We neglect the contribution from the reproduction and  mortality of the predators, assuming that these processes are much more slow than the other processes considered. The prey spreading is purely diffusive but the predators spreading is resulted from  both diffusion and advection.    Equation (\ref{TxEq})  describes the evolution of the macroscopic velocity  of the predators in responce to the prey density  and  to the external signal denoted as  $f$; parameter $\kappa$ measures the prey-taxis intensity.  We also take into account the diffusion of the predators velocity and the resistance to their motion due to the environment and denote the corresponding coefficients as $\delta_u$ and $\nu$, respectively.

The homogeneous version of system (\ref{TxEq})--(\ref{PryEq}) (in which $f=0$) had been proposed in \cite{GMT,AGMTS} as a simple model of inertial prey-taxis. Indeed, the flux of the predators has to overcome certain inertia while taking the direction of  the stimulus gradient  unlike what is presumed by the PKS transport model. This can be seen from equation (\ref{TxEq}).  Nevertheless, the integrating of Eq.~(\ref{TxEq}) with the use of velocity potential puts the  system (\ref{TxEq})-(\ref{PryEq}) back into the  PKS  framework \cite{TtnZgr}. Namely, ansatz  $u=\kappa\partial_x\phi$  transforms  system (\ref{TxEq})-(\ref{PryEq}) into the system consisting of equation (\ref{PryEq}) and of the following two equations
\begin{equation}\label{PKSflxEq}
\pt_t\phi=q +\kappa^{-1}f-\nu \phi+\delta_u \pt^2_x\phi;\  \pt_t p=\pt_x(\delta_p\pt_xp-\kappa p\pt_x\phi) \
\end{equation}
The second equation here has got  PKS term which is proportional to $\kappa$. Hence $\kappa$ can be seen as  measure of the predators motility in response to a signal the role of which is played by the velocity potential, $\phi$. The first equation in (\ref{PKSflxEq}) describes the  signal production. Note that the stimulus for taxis is not the prey itself but the signal emitted by it.  The interactions like this are classified as  indirect prey-taxis (see e.g. \cite{Ts94}-\cite{Ts04-1}, \cite{TllWrzk}, \cite{TtnZgr},\cite{LiTao}).
\section{Homogeneous environment}
\label{SecHmgEnv}
\setcounter{equation}{0}
\noindent
%The homogeneous version of model (\ref{TxEq})-(\ref{PryEq}) had been investigated  in \cite{GMT,AGMTS}. One of key findings of these studies is that a sufficiently strong prey-taxis makes the system capable of replacing the homogeneous equilibria by  wavy motions in response to the increase in the prey deficiency. The waves occurs due to the Poincare-Andronov-Hopf (PAH) bifurcation. For the sake of completeness, we briefly describe these results here.

The homogeneous system (\ref{TxEq})-(\ref{PryEq})  has a family of homogeneous equilibria in which
\begin{equation}\label{Eqlbr}
p\equiv p_e,\ q\equiv q_e,\ u=0,\  p_e=\const>0,  q_e=\const>0,\ p_e+q_e=1.
\end{equation}
The linearization of homogeneous system (\ref{TxEq})-(\ref{PryEq}) nearby an  equilibrium of family (\ref{Eqlbr}) specified with density $p_e$ takes the form
\begin{eqnarray}
&\pt_t {u}+\nu  {u}- {\kappa}\pt_{x} {q}=\delta_u\pt^2_{x}u,&
\label{TxEqLnr}\\
& \pt_t{p}+{p}_e \pt_{x} {u}=\delta_p \pt_{x}^2 {q};&
\label{PrdEqLnr}\\
& \pt_t {q}+q_e( {p}+ {q})=\delta_q \pt_{x}^2 {q};&
\label{PryEqLnr}\\
&  {p}_e+ {q}_e=1. &
\nonumber
\end{eqnarray}
The   eigenmodes of  small perturbations of equilibria (\ref{Eqlbr}) have  the following form
\begin{equation}\label{EgnMds}
       (\hat{u},\hat{p},\hat{q})\exp(i\alpha x+\lambda t),\ \lambda  \in \mathbb{C},\ \alpha\in \mathbb{R}.
\end{equation}
Here $\lambda$ is the  eigenvalue of the  spectral problem arising from  the substituting of eigenmodes (\ref{EgnMds}) into linear system (\ref{TxEqLnr})-(\ref{PryEqLnr}).  We say that  eigenmode (\ref{EgnMds}) is stable  (unstable, neutral) if the real part of $\lambda$ is negative (positive, equal to zero). We'll be looking for the occurrences of instability, that is, the transversal intersections of the imaginary axis by a smooth branch of eigenvalues, $\lambda$,  when the other parameters of the spectral problem  change themselves along a smooth path. If such a branch crosses  the imaginary axis at a non-zero point, the instability is named  as  oscillatory, or, otherwise, it is  named as  monotone.

It is well-known that an occurrence of instability in the family of equilibria is necessary for the local bifurcations. If there are no additional degenerations the monotone instabilities are accompanied with branching of the equilibria  family,   the oscillatory  instabilities are accompanied with the limit cycle  branching off  from the basic family (Poincare-Andronov-Hopf  bifurcation), and more complex  bifurcations happen  in the case of additional degeneracy e.g.  when neutral spectrum  is multiple.

%To be more specific, we present here the underlying results regarding the eigenmodes  of equilibria  (\ref{Eqlbr}). The  details  are  in Appendix I.
%Here the normalized `total mass' of the predators, $p_e$,  plays the role of the family parameter. The oscillatory instability  occurs in family (\ref{Eqlbr}) in response to the increase in $p_e$ provided that parameter $\kappa$ in Eq.~(\ref{TxEq}) is greater than a threshold.

It is convenient to introduce the following notation
$$
\beta=\alpha^2,\ \delta=(\nu,\delta_q,\delta_p,\delta_u).
$$
Let us cut off excessive degeneracy by assuming the following
\begin{equation}\label{Restr}
\beta>0,\ 0<p_e<1,\    \nu(\delta_p+\delta_u+\delta_q)>0.
\end{equation}
Note that each equilibrium (\ref{Eqlbr}) has a neutral homogeneous mode (that corresponds to $\lambda=\alpha=0$)  but  this does not lead to any long-wave instabilities.
%Since the homogeneous system (\ref{TxEq})-(\ref{PryEq}) is invariant with respect to the reflections $x\to-x$, $u\to -u$, the spectrum of eigenmodes is invariant with respect to the reflections $\alpha\to-\alpha$. Therefore we consider the  positive wave numbers only.

\begin{figure}[h]
\centering
\includegraphics [scale=0.50]{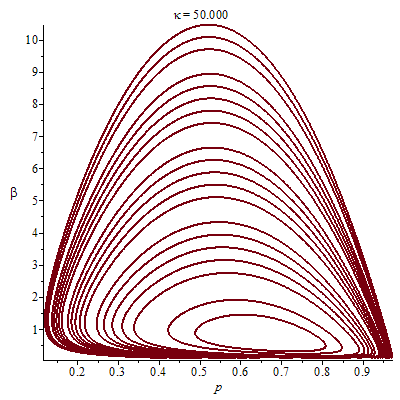}\hfill\includegraphics [scale=0.50]{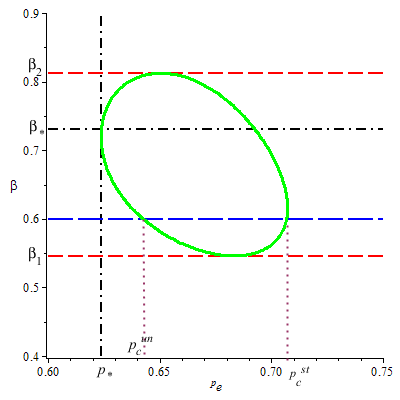}
\caption{\small The {\sl left frame} shows typical curves determined by equation  $\kappa_c(p_e,\beta,\delta)=\kappa$ on the $(p_e,\beta)$ plane for $\kappa$  in the range from $15.12$ to $50.0$  and $\delta=(1,1,0,0)$. The area bounded by a curve increases with $\kappa$. The {\sl right frame} corresponds to $\delta=(1,1,0,0)$ and
$\kappa=13.56$. The other numerical values are: $p_*\approx 0.624$; $p_c^{un}\approx0.642$, $p^{st}_c\approx0.707$, $\beta_1\approx0.546$,  $\beta_2\approx0.812$}
\label{fig1}
\end{figure}

Let $\Pi$ be a domain cut out by  inequalities (\ref{Restr}) in the space of parameters $p_e,\beta,\delta$.  Let us consider  an equilibrium of family (\ref{Eqlbr}) with specific density $p_e$ and its  eigenmodes  (\ref{EgnMds}) with specific wave length $\alpha=\sqrt{\beta}$.  There exists function
$$
\kappa_c = \kappa_c (p_e, \beta,\delta)
$$
analytic in $\Pi$ and such that (i) each of those  eigenmodes  is stable provided that $\kappa<\kappa_c (p_e, \beta,\delta)$; (ii) there is an unstable mode  provided that $\kappa>\kappa_c (p_e, \beta,\delta)$; (iii) there exists two conjugated neutral modes with $\lambda\neq 0$ provided that $\kappa=\kappa_c(p_e, \beta,\delta)$. The oscillatory instability  occurs  each time  a path in  $\Pi\times(0,\infty)$ (where $(0,\infty)\ni\kappa$) intersects  graph $\{(p_e,\beta,\delta),\kappa_c(p_e,\beta,\delta)\}$  transversally, perhaps, except for some cases of degeneracy (Fig.~\ref{fig1}).  Note that
\begin{equation}\label{DfTrhKpp} \min\limits_{0<p_e<1,\beta>0}\kappa_c(p_e,\beta,\delta)=\kappa_*(\delta)>0
\end{equation}
where the strict positiveness  takes place for every $\delta$ obeying (\ref{Restr}). For every $\kappa=\kappa_*(\delta)$, equation $\kappa=\kappa_c(p_e,\beta,\delta)$ determines a closed curve inside semistrip $\{0<p_e<1,\beta>0\}$ which  tends to the boundary of the semi-strip as $\kappa\to+\infty$.
Fig.~\ref{fig1} shows that the oscillatory instability of homogeneous equilibria occurs in response to growth of the predators density, $p_e$, provided that their motility is great enough; that is, $\kappa>\kappa_*(\delta)$.  It follows from the results of \cite{GMT,{AGMTS}}\footnote{In \cite{GMT,{AGMTS}},   a bounded spatial domain is considered and the corresponding spectra  of eigenmodes (\ref{EgnMds}) fill certain discrete subsets of the continuous spectra described above.} that this oscillatory instability is accompanied by Poincare-Andronov-Hopf bifurcation manifested   by  excitation of  waves, and, moreover,   the  wave dynamics turns out to be  more advantageous than equilibrium in the sense that predators can consume more, while leaving greater stock of the prey. The value of $\kappa_* (\delta) > 0$ is the threshold for the predators motility, $\kappa$. If $\kappa<\kappa_* (\delta)$ then neither the oscillatory instability of the homogeneous equilibrium nor the accompanying bifurcation are possible, whatever  values of the predator density and the disturbances wavelengths  are specified. Thus, the weakening of the predators motility leads to the absolute stabilization of the homogeneous equilibria and does not allow the community  to adapt itself to the resource deficiency.

The Poincare-Andronov-Hopf bifurcation  has been studied for more general classes of the PKS type systems in \cite{Chaplain,TtnZgr,QiJiLu}.
\section{Fluctuating environment}
\label{SecFlctEnvr}
\setcounter{equation}{0}
\noindent
In what follows, let's consider every function of variables $(\xi,\tau)$ as a function defined on 2-torus $\mathbb{T}^2$. Define
\begin{equation}\label{DfnAvr}
    \langle g\rangle=\frac{1}{4\pi}\int\limits_{0}^{2\pi}\int\limits_{0}^{2\pi} g(x,t,\xi,\eta)\,d\xi d\eta.
\end{equation}
Let the external signal in  Eq.~(\ref{TxEq}) be a short wave, i.e.
\begin{equation}\label{ShrtWvSnl}
 f=f(x,t,\xi,\tau),\ \xi=\omega x,\ \tau=\omega t,\ \omega\gg 1
\end{equation}
and let the diffusion rates in Eq.~(\ref{TxEq}-\ref{PrdEq})  be of the same order as the  wave length, namely:
\begin{equation}\label{Dff2bSmll}
\delta_u=\nu_1\omega^{-1},\ \delta_p=\nu_2\omega^{-1},\ \nu_1>0,\ \nu_2>0.
\end{equation}
With these assumptions,  asymptotic of  system (\ref{TxEq})-(\ref{PryEq})  takes the following form
\begin{eqnarray}
& q(x,t)=\bar{q}(x,t)+O({\omega}^{-1}),\ \omega\to+\infty;&
\label{q=}\\
&u(x,t)=\bar{u}(x,t)+\widetilde{u}_0(x,t,\tau,\xi)+O({\omega}^{-1}),\  \omega\to+\infty;&
\label{u=}\\
& p(x,t)=\bar{p}(x,t)P(x,t,\tau,\xi)+O({\omega}^{-1}),
\ \omega\to+\infty &
\label{p=}\\
&\partial_\tau \widetilde{u}_0=\partial_\xi (f +\nu_1\partial_\xi \widetilde{u}_0);\  \langle  \widetilde{u}_0\rangle=0,&
\label{Tldu0Eq}\\
&\partial_\tau P=\partial_\xi(\nu_2\partial_{\xi}P -P(\bar{u}+\widetilde{u}_0)),\ \langle  P\rangle=1;&
\label{PEq}\\
& \partial_t\bar{u}=\partial_x(\kappa \bar{q}+\sigma \bar{f})-\nu \bar{u};\ \bar f=\langle f \rangle &
\label{EqUSlw0}\\
 & \partial_t \bar{p}+\partial_x(\bar{p}(\bar{u}+\langle \widetilde{u}_0P\rangle))=0;&
\label{EqPSlw0}\\
&\bar{q}_t=\bar{q}(1-\bar{p}-\bar{q})+\delta_q\partial^2_x\bar{q}.&
\label{EqQSlw0}
\end{eqnarray}
Note that
$
\bar {u}=\langle u\rangle+O({\omega}^{-1}),\quad \bar {p}=\langle p\rangle+O({\omega}^{-1}),\quad \bar {q}=\langle q\rangle+O({\omega}^{-1}),\quad \omega\to+\infty.
$

Eq.~(\ref{EqPSlw0}) governing the predators transport describes, in particular, a drift   the velocity of which is $\langle \widetilde{u}_0P\rangle$. This drift  is the only remembrance of the short-wavelength signal. The drift velocity is uniquely determined by $\bar{u}$ provided $f$ is specified. Therefore, Eqs.~ (\ref{EqUSlw0}), (\ref{EqPSlw0}) and (\ref{EqQSlw0}) form a closed system relative to unknowns $\bar{p},\bar{q},\bar{u}$. This is what we call as the homogenized system.

The details of derivation are placed in Appendix II.
\section{Stabilization of quasi-equilibria}
\label{SecRltvEql}
\setcounter{equation}{0}
\noindent
In what follows,  we assume that
\begin{equation}\label{FOnXiTau}
    f={f}(\xi,\tau).
\end{equation}
Under this assumption, Eqs.~(\ref{Tldu0Eq})-(\ref{PEq}) imply that $\widetilde{u}_0$ is independent of $(x,t)$, and $P$ is independent of $(x,t)$ provided $\bar{u}=0$. Hence the homogenized system  has a family of homogeneous equilibria
\begin{equation}\label{REqlbr}
\bar{p}\equiv p_e,\ \bar{q}\equiv q_e,\ \bar{ u}=0,\  p_e=\const>0,  q_e=\const>0,\ p_e+q_e=1.
\end{equation}
Formulae (\ref{q=})--(\ref{p=}) together with  Eqs.~(\ref{Tldu0Eq})-(\ref{PEq})  identify every equilibrium (\ref{REqlbr}) with a short-wavelength solution of the original system which we call \emph{quasi-equilibrium}. In the quasi-equilibria, the predators velocity vanishes on average while the averaged densities of both species are  constant (at least, in the leading approximation).  In the considerations below the quasi-equilibria of the original system and homogeneous equilibria of the homogenized system are not actually distinguishable,  and   we shall use the same name for both.

Let us examine the linear stability of quasi-equilibria.
Let  $f=f(\xi,\tau)$ be given and $\widetilde{u}_0=\widetilde{u}_0(\xi,\tau)$ be determined by Eq.~(\ref{Tldu0Eq}).
Define the mapping $\mathbb{R}\to \mathbb{R}$ as follows
 \begin{equation}\label{VActsOnR}
    \eta\mapsto \langle\widetilde{u}_0 P\rangle
\end{equation}
where $P=P(\xi,\tau)$  is the solution to problem
\begin{equation}\label{EtaToP}
    \partial_\tau P=\partial_\xi(\nu_2\partial_{\xi}P -P(\eta+\widetilde{u}_0)),\ \langle  P\rangle=1.
\end{equation}
Let $\mathcal{V}^\prime({f})$  denote   the differential of mapping (\ref{VActsOnR})-(\ref{EtaToP}) evaluated at the origin.
%where $\mathcal{V}$  is  determined by   (\ref{DfV}).
The system governing  the evolution of a small perturbation of a given quasi-equilibrium takes the form
\begin{eqnarray}
& \pt_t\bar{u}+\nu \bar{u}-\kappa  \pt_{x}\bar{q}=0;&
\label{TxEqLnrSlw}\\
& \pt_t\bar{p}+ {p}_e(1+\mathcal{V}^\prime({f}))\pt_{x}\bar{u}=0;&
\label{PrdEqLnrSlw}\\
& \pt_t\bar{q}+{q}_e(\bar{p}+\bar{q})-\delta_q \pt^2_{x}\bar{q}=0;&
\label{PryEqLnrSlw}
\end{eqnarray}
Note that factor $1+\mathcal{V}^\prime({f})$ can be treated as  the effective motility coefficient.

Re-scaling $\bar u$, we transform   system (\ref{TxEqLnrSlw})-(\ref{PryEqLnrSlw}) into the form (\ref{TxEqLnr})-(\ref{PryEqLnr}) with $\delta_p=\delta_u=0$, and $\kappa$  changed by $\bar\kappa=(1+\mathcal{V}^\prime({f}))\kappa$. Therefore,  the effect of the  short-wavelength external signal on  the stability of quasi-equilibria manifests itself in altering the prey-taxis intensity or, equivalently, the predators motility in accordance with rule $\kappa\mapsto (1+\mathcal{V}^\prime({f}))\kappa$.

Note  that  restrictions (\ref{Restr}) on the problem parameters imposed in Sec.~\ref{SecHmgEnv} allows $\delta_p$ and $\delta_u$ to be zero simultaneously. Therefore all the statements about the   stability of equilibria of the homogeneous version of system (\ref{TxEq})-(\ref{PryEq}), made in Sec.~\ref{SecHmgEnv}, are also true regarding the stability of the quasi-equilibria, provided that  the predators motility, $\kappa$, is replaced by its effective counterpart $\bar{\kappa}=(1+\mathcal{V}^\prime({f}))\kappa$. In particular, the inequality
\begin{equation}\label{NewAbsTrhld0}
1+\mathcal{V}^\prime({f})<{\kappa}^{-1}{\kappa_*|_{\delta_p=\delta_u=0}}
\end{equation}
implies absolute stabilization of quasi-equilibria in the sense that there is no instability irrespective of what equilibrium and what the perturbation wave number are considered (here $\kappa_*$ is exactly the threshold motility of the predators introduced in Sec.~\ref{SecHmgEnv}).

The righthand side in inequality (\ref{NewAbsTrhld0})  does not depend on  the external signal while the  lefthand side  evidently does depend and it is  interesting to learn to what extent; in particular, whether or not a  short-wavelength external signal is capable of the absolute stabilizing the equilibria which are unstable provided that  the signal is off. More formally, the question is whether or not one can get  inequality (\ref{NewAbsTrhld0})  by manipulating function $f$. Let's show  that the  answer to this question is affirmative.

Let us denote as  $\langle g \rangle^\xi$ the average value of some function $g=g(x,t,\xi,\tau)$  with respect to variable $\xi$; that is,
\begin{equation}\label{XiAvrg}
    \langle g \rangle^\xi=\frac{1}{2\pi}\int\limits_0^{2\pi}g(x,t,\xi,\tau)\,d\xi
\end{equation}
Let the external signal be independent of $\tau$:
\begin{equation}\label{FonXi}
{f}={f}(\xi),\  \langle{f}\rangle=0.
\end{equation}
Under condition (\ref{FonXi}), evaluation of  $\tilde{u}_0$ reduces to the solving  of  equations
\begin{equation}\label{TldU0OnXi}
\partial_\xi (\sigma f +\nu_1\partial_\xi \widetilde{u}_0)=0,\ \langle u_0\rangle^\xi=0.
\end{equation}
Then  evaluation of $P$ reduces to the solving  of  equations
\begin{equation}\label{PonXi}
    \partial_\xi(\nu_2\partial_{\xi}P -P(\bar{u}+\widetilde{u}_0))=0,\ \langle  P\rangle^\xi=1,
\end{equation}
 Consequently,
\begin{equation}\label{DffV}
    \mathcal{V}^\prime(f)=\langle  \widetilde{u}_0P_1\rangle^\xi
\end{equation}
where $P_1$ is determined by equations
\begin{eqnarray}
   & \partial_\xi(\nu_2\partial_{\xi}P_1 -P_0-\widetilde{u}_0P_1)=0,\ \langle  P_1\rangle^\xi=0,&
 \label{P1onXi}  \\
 & \partial_\xi(\nu_2\partial_{\xi}P_0-\widetilde{u}_0P_0)=0,\ \langle  P_0\rangle^\xi=1.
 \label{P0onXi}
\end{eqnarray}
%The integrating of differential equations of system (\ref{P1})-(\ref{P0})
%\begin{eqnarray}
%&\nu_2\partial_\xi P_{0}=\tilde{u}P_0-C_0,\quad \langle P_0\rangle=1&
%\label{P0Eq}\\
%&\nu_2\partial_\xi P_{1}=\tilde{u}P_1+P_0 -C_1,\quad \langle P_1\rangle=0 &
%\end{eqnarray}
%where $C_0,C_1$ are the constants of integration.
Here we are interested  only in those   solutions to  equations (\ref{TldU0OnXi}), (\ref{PonXi}), (\ref{P1onXi}) and (\ref{P0onXi}) which are $2\pi-$periodic in variable $\xi$.

Straightforward calculations, taking into account the periodicity, yield the following results
\begin{eqnarray}
&P_0=A_0\mathrm{e}^v,\quad A_0=\frac{1}{\langle\mathrm{e}^v{\rangle^{\xi}}};&
\label{P0=}\\
&P_1=\mathrm{e}^v(A_1+\frac{\partial_\xi^{-1}(A_0-C_1\mathrm{e}^{-v})}{\nu_2}),\  A_1=\frac{A_0\langle\mathrm{e}^v\partial_\xi^{-1}(C_1\mathrm{e}^{-v}-A_0){\rangle^{\xi}}}{\nu_2},\  C_1=\frac{1}{\langle\mathrm{e}^{-v}{\rangle^{\xi}}\langle\mathrm{e}^v{\rangle^{\xi}}};&
\label{P1=}\\
&v=(\nu_2\partial_\xi)^{-1} \tilde{u}_0,&
\label{v=}
\end{eqnarray}
where transformation $\partial_\xi^{-1}$ acts on $2\pi$-periodic (in $\xi$) functions vanishing on average as the  right inverse to  $\partial_\xi$, i.e.
$$
\partial_\xi \partial_\xi^{-1} w=w,\ \langle \partial_\xi^{-1} w\rangle^\xi=0
$$
for any function  $w$ such that it is $2\pi$-periodic in $\xi$ and $\langle w\rangle^\xi=0$.
Consequently,
$$
 \mathcal{V}^\prime({f})= \nu_2\langle (A_1+\nu_2^{-1}\partial_\xi^{-1}(A_0-C_1\mathrm{e}^{-v}))\mathrm{e}^v\partial_\xi v{\rangle^{\xi}}=-\langle \mathrm{e}^v(A_0-C_1\mathrm{e}^{-v}){\rangle^{\xi}}=C_1-1=
 \frac{1}{\langle\mathrm{e}^{-v}{\rangle^{\xi}}\langle\mathrm{e}^v{\rangle^{\xi}}}-1.
$$
Finally, the solving of  problem (\ref{TldU0OnXi}) yields $\widetilde{u}_0= - (\nu_1\partial_\xi)^{-1}  {f}$, and then we get
$$
v=(\nu_2\partial_\xi)^{-1} \widetilde{u}_0=-(\nu_2\partial_\xi)^{-1}(\nu_1\partial_\xi)^{-1}  {f}.
$$
In what follows we change $f$ by $-f$. This does not lead to any mistakes since functional  $\mathcal{V}^\prime({f})$ is even with respect to $f$.
%\begin{equation}\label{V'=1}
  %  1+\mathcal{V}^\prime({f})=
   % \frac{1}{\langle\mathrm{e}^{-v}{\rangle^{\xi}}\langle\mathrm{e}^{v}{\rangle^{\xi}}},\quad v=-\sigma(\nu_1\nu_2)^{-1}\partial_\xi^{-2}{f},
%\end{equation}
Thus the effective motility $\bar\kappa= (1+\mathcal{V}^\prime({f}))\kappa$ takes almost explicit form; namely
\begin{equation}\label{EffMtlt}
    \bar\kappa=\frac{\kappa}{\langle\mathrm{e}^{-v}{\rangle^{\xi}}\langle\mathrm{e}^{v}{\rangle^{\xi}}},\quad v=(\nu_1\nu_2)^{-1}\partial_\xi^{-2}{f}.
\end{equation}
Substituting this into inequality (\ref{NewAbsTrhld0}) yields a more explicit form of the criterion for absolute stabilization:
\begin{equation}\label{NewAbsTrhld}
\frac{1}{\langle\mathrm{e}^{-v}{\rangle^{\xi}}\langle\mathrm{e}^{v}{\rangle^{\xi}}}<\kappa^{-1}\kappa_*|_{\delta_p=\delta_u=0},
\end{equation}
\begin{Exmpl}
\label{Exmplsn}
\emph{Let $f=A\sin\xi$, $A=\const>0$. Then $v=-A(\nu_1\nu_2)^{-1}\sin\xi$, and
$$
\kappa\bar{\kappa}^{-1}=\langle\mathrm{e}^{-v}{\rangle^{\xi}}\langle\mathrm{e}^{v}{\rangle^{\xi}}=I^2_0(a),\ a=\frac{A}{\nu_1\nu_2}
$$
where $I_0$ is the modified Bessel function of first kind. Consequently,  the criterion for absolute  stabilization (\ref{NewAbsTrhld}) takes the following form
\begin{equation}\label{NewAbsTrhldSn}
    I^{-2}_0(a)<{\kappa}^{-1}\kappa_*|_{\delta_p=\delta_u=0},\ a=\frac{A}{\nu_1\nu_2}
\end{equation}
The left hand side in this inequality decreases exponentially when parameter $a$ grows up. This parameter  represents a characteristic amplitude of the external signal. Hence, the increase in the level of the external signal leads to absolute stabilization of the relative equilibria and this effect is rather powerful in the sense that the threshold of the  absolute stabilization is attained exponentially fast.}
\end{Exmpl}
\textbf{Remark 1.}
\label{Rmrk2} In fact,  the  exponential decrease in  effective motility of the predators in response to the increase  in the level of  external signal is  a generic property of the system.
This can be seen by estimating expression (\ref{EffMtlt}) using the Laplace method for $f=Af_0$, where $A\to\infty$ while choice of $f_0$ is rather wide.
 %for  estimation of $\langle\mathrm{e}^{v}{\rangle^{\xi}}$.
Therefore, \emph{the increase in the amplitude of short-wavelength external signal typically stabilizes  the relative equilibria as is shown in Example~\ref{Exmplsn}.}
\smallskip\\
\textbf{Remark 2.}
\label{Rmrk3}
The stabilization described above occurs irrespective of whether the external signal is attractive or repellent for the predators.

\noindent
\section{Conclusions}
\noindent
Thus the short-wavelength signal applied to PKS-type equations is capable of reducing the cross-diffusive transport on average drastically.  Typically, the decrease  in cross-diffusive motility in response to the increase in the signal level is exponential, and such the loss of motility exerts very powerful stabilizing effect on the primitive quasi-equilibria. An interesting question is to what extent the shape of the external  signal is able to enhance or weaken  the stabilizing effect of it. This question leads to an optimization problem for the effective motility defined in (\ref{EffMtlt}) subject to restriction
$
\langle f^2\rangle=1.
$

Finally, we would like to relate the results described above to general phenomena which are often observed upon the application of the averaging and homogenization techniques.  First, the additional cross-diffusive flux in Eq.~(\ref{EqPSlw0}) definitely resembles Stokes's drift which typically arises upon the averaging of  the advection  of  some matter   over small-scale oscillations of the advective velocity \cite{Vldmrv1}. Second,   the short-wave stabilizing of quasi-equilibria resembles similar effects of high-frequency vibrations on the pendulum dynamics  or on the stratified fluid dynamics \cite{Yudovich,Vldmrv}.
\subsubsection*{Acknowledgments} Andrey Morgulis acknowledges the support from Southern Federal University (SFedU)
%\subsubsection*{Data accessibility statement}
%This work does not have any experimental data.
%\subsubsection*{Competing interests statement}
%We have no competing interests.
%\subsubsection*{Authors’ contributions.} All of the results presented in this article equally belongs to both authors. All authors gave final approval for publication.
%\subsubsection*{Funding.}
%This work was supported by Southern Federal University (SFedU).

\section{Appendix I. Eigenmodes  of the equilibria }
\setcounter{equation}{0}
\noindent
Let us choose an equilibrium out of family (\ref{Eqlbr}) by specifying a value of the family parameter, $p_e$.
% The   system governing small perturbation of this equilibrium  takes the form
%\begin{eqnarray}
%&  {u}_t+\nu {u}-{\kappa} {q}_x=\delta_u u_{xx}&
%\label{TxEqLnr}\\
%&  {p}_t+ p_e u_x=\delta_p p_{xx}&
%\label{PrdEqLnr}\\
%&  {q}_t+q_e(p+q)=\delta_q q_{xx}&
%\label{PryEqLnr}
%\\
%&\ p_e+q_e=1.
%\nonumber
%\end{eqnarray}
%The   system derived from  the linearization of the homogeneous version of system (\ref{TxEq})-(\ref{PryEq}) near a ground state (\ref{Eqlbr}) specified with certain $p_e\in (0,1)$ is as follows
%\begin{eqnarray}
%&  {u}_t+\nu {u}-{\kappa} {q}_x=\delta_u u_{xx}&
%\label{TxEqLnr}\\
%&  {p}_t+ p_e u_x=\delta_p p_{xx}&
%\label{PrdEqLnr}\\
%&  {q}_t+q_e(p+q)=\delta_q q_{xx}&
%\label{PryEqLnr}
%\end{eqnarray}
Eigenvalues for the eigenmodes (\ref{EgnMds}) having a specific  wavenumber $\alpha$ are solutions to the following algebraic equation
\begin{eqnarray}
&\lambda^3+(D_1+D_2+D_3)\lambda^2+(D_2D_3+D_1D_3+D_1D_2)\lambda +D_1D_2D_3+\beta\kappa p_e q_e=0,
& \label{ChrEq} \\
&D_1 = \nu+\beta\delta_u;\ D_2=\beta\delta_p;\ D_3=q_e+\beta\delta_q;\ \beta=\alpha^2.&
\nonumber
\end{eqnarray}
In view of  restrictions (\ref{Restr}),  all  the coefficients of the polynomial on the left hand side of Eq.~(\ref{ChrEq}) are strictly positive. Consequently, this polynomial has  neither  positive nor zero roots. Hence, no unstable or neutral eigenmode corresponds to a real eigenvalue.

It follows from the Routh-Hurwitz theorem that the necessary and sufficient condition for all roots of polynomial (\ref{ChrEq})
to be in the open  left half-plane of the complex plane is
$$
(D_1+D_2+D_3)(D_2D_3+D_1D_3+D_1D_2)>D_1D_2D_3+\beta\kappa p_e q_e.
$$
This inequality admits a more compact form, namely:
$$
(D_1+D_2)(D_1+D_3)(D_2+D_3)>\beta\kappa p_e q_e.
$$
Hence the results of linear analysis described in  Sec.~\ref{SecHmgEnv} are valid provided that  one defines  the threshold magnitude for the predators motility  as
\begin{equation}\label{KppaCr}
   \kappa_c=\frac{(D_1+D_2)(D_1+D_3)(D_2+D_3)}{\beta p_e q_e}.
\end{equation}
Indeed, the degree of  polynomial (\ref{ChrEq}) is 3. Consequently, $\kappa=\kappa_c$ is the necessary and sufficient  condition for a root of polynomial (\ref{ChrEq}) to lie on
the imaginary axis (this root cannot be zero).

%Let us consider the threshold value $\kappa_*$ that has been defined in  (\ref{DfTrhKpp}). Although we do not have an explicit expression for $k_*$, its positiveness as well as the other properties mentioned in Sec. ~\ref{SecHmgEnv} are almost obvious or, in any case, can be verified directly using
%expression (\ref{KppaCr}).   A useful auxiliary result is the explicit   minimization of $\kappa_c$ relative to  $q_e$:
%\begin{eqnarray*}
  %&\min\limits_{0<q_e<1}\kappa_c=\frac{c}{\beta}
%\left(\frac{\sqrt{a(1+a)}+\sqrt{b(1+b)}}{\sqrt{(1+a)(1+b)}-\sqrt{ab}}\right)^2\quad\ \text{where} &  \\
%  &a=(\delta_{p}+\delta_{q})\beta,\ b=\nu+\beta(\delta_{u}+\delta_{q}),\ c=\nu+\beta(\delta_{p}+\delta_{u}). &
%\end{eqnarray*}
In the case $\delta_u=\delta_p=0$ which is of particular interest for the considerations of Sec.~\ref{SecRltvEql}, the threshold    takes  much simpler form
$$
\kappa_c|
    _{\delta_p=\delta_u=0}={\frac {{  \nu}\, \left( {  \delta_q}\,\beta+{  \nu}+q_e \right)  \left( {
  \delta_q}\,\beta+q_e \right) }{q_e\, p_e \beta}}.
$$
\section{Appendix II. Derivation of the asymptotic }
\setcounter{equation}{0}
\noindent
%Let the external signal be specified as $f=f(x,t,\xi,\tau)$ where $\xi=\omega x$ and $\tau=\omega t$ in accordance  with assumption
%(\ref{ShrtWvSnl}).  Let the diffusion rates satisfy condition (\ref{Dff2bSmll}).  For such a short-wavelength signal, we seek a short-wavelength response depending
%on variables $(x,t,\xi,\tau)$. In what follows, we assume that both the external signal and the system response are $2\pi-$periodic in $\xi=\omega x$, $\tau=\omega t$.
Here we derive formally the asymptotic approximation described in Sec.~\ref{SecFlctEnvr}.

On introducing the fast variables $\xi=\omega x$, $\tau=\omega t$ into the governing equations (\ref{TxEq})-(\ref{PryEq}), they take the following form
\begin{eqnarray}
  &\omega((\partial_t+\omega\partial_\tau)u- (\partial_x+\omega\partial_\xi) (\kappa q+\sigma f)+\nu u)=\nu_1 (\partial_x+\omega\partial_\xi)^2u;&
  \label{TxEq2Scl}\\
  &\omega((\partial_t+\omega\partial_\tau)p+(\partial_x+\omega\partial_\xi)(up))=\nu_2 (\partial_x+\omega\partial_\xi)^2p;&
  \label{PrdEq2Scl}\\
  &(\partial_t+\omega\partial_\tau)q-q(1-p-q)=\delta_q (\partial_x+\omega\partial_\xi)^2q.&
  \label{PryEq2Scl}
\end{eqnarray}
 We look for an asymptotic  expansion of the solution to system (\ref{TxEq2Scl})-(\ref{PryEq2Scl}) in the form
 \begin{equation}\label{AsmptSrs}
    (u,p,q)=\sum_{k\geqslant 0}\omega^{-k}(u_k,p_k,q_k)(x,t,\xi,\tau), \omega\to\infty.
\end{equation}
where all coefficients are required to be  $2\pi-$periodic in $\xi$ and $\tau$.  Substitution of (\ref{AsmptSrs}) into
 system (\ref{TxEq2Scl})-(\ref{PryEq2Scl}) and collecting terms of equal order in $\omega$ yields a sequence of equations
 which will be solved step by step.

Terms of order $\omega^2$ in Eq.~(\ref{PryEq2Scl}) lead to the equation
\begin{equation}\label{EqQ0}
  \partial_{\xi\xi}q_{0}=0
\end{equation}
that obviously has no periodic  solutions  except for solutions independent of $\xi$.  Thus
\begin{equation}\label{q0=}
q_0={q}_0(x,t,\tau).
\end{equation}
where function ${q}_0$ is to be determined at the subsequent steps.
In view of (\ref{q0=}),  the collecting of terms of order $\omega^2$ in Eqs. (\ref{TxEq2Scl}-\ref{PrdEq2Scl}) leads to the equations
\begin{eqnarray}
& (\partial_\tau-\nu_1\partial_{\xi\xi})u_0=\sigma\partial_\xi f; &
\label{EqU0}\\
&(\partial_\tau-\nu_2\partial_{\xi\xi})p_0+\partial_\xi(u_0p_0)= 0;&
\label{EqP0}\\
&(u_0,p_0)=(u_0,p_0)(x,t,\xi,\tau),\ (\xi,\tau)\in \mathbb{T}^2.&
\nonumber
\end{eqnarray}
Note that equation (\ref{EqU0}) is exactly the first equation in problem (\ref{Tldu0Eq}). Equation (\ref{EqU0}) has only one periodic solution vanishing on average in the sense of definition (\ref{DfnAvr}). We denote this solution as $\widetilde{u}_0$. Thus  $u_0=\bar{u}+\widetilde{u}_0$, $\bar u=\langle u_0\rangle$, and we have justified the leading term in the  asymptotic approximation for $u$, given by  (\ref{u=}), (\ref{Tldu0Eq}).

We  need the following
\\
\textbf{Lemma.} {\sl Let $w=w(\xi,\tau)$ be a smooth $2\pi$-periodic (in both $\xi$ and $\tau$) function. Consider equation}
\begin{equation}\label{QEq}
\partial_\tau Q+\partial_\xi(wQ-\eps\partial_{\xi}Q)=0\ \text{on}\ \mathbb{T}^2;\ \eps=\const>0.
\end{equation}
{\sl Then there exists a unique $2\pi$-periodic (in $\xi$ and $\tau$)  solution to Eq.~(\ref{QEq}) satisfying the additional condition}
\begin{equation}\label{QAvr}
    \langle Q\rangle=1.
\end{equation}
Proof of this statement will be given in the end of this Appendix.

We continue constructing the asymptotic expansion.
Equation~(\ref{EqP0}) coincides with (\ref{QEq}) up to the replacing $\eps$ by $\nu_2$ and $w$ by $\bar{u}+\widetilde{u}_0$, so that we can apply the above lemma, which yields
\begin{equation}\label{p0=}
    p_0=\bar{p}(x,t)P(x,t,\xi,\tau),
\end{equation}
where $P$ is uniquely determined by (\ref{PEq}). Hence we have justified the the leading term in the asymptotic  approximation  determined by (\ref{p=}) for unknown $p$.

Now let us consider the terms of order $\omega$.  From Eq.~(\ref{PryEq2Scl}), with the help of (\ref{q0=}), we get
\begin{equation}\label{EqQ1}
    \delta_q\partial_{\xi\xi}q_1=\partial_\tau q_0,
\end{equation}
where $q_0$ does not depend on $\xi$. Hence, this equation has a periodic solution if and only if $q_0$ does not depend on $\tau$. Thus $q_0=\bar{q}(x,t)$ and we arrive at the leading term in the asymptotic approximation (\ref{q=}). Further,  every  solution to Eq.~(\ref{EqQ1}) has to have a form
\begin{equation}\label{q1=}
    q_1=q_{1}(x,t,\tau).
\end{equation}
Functions $q_1$ and $\bar{q}$ are to be determined at  subsequent steps of the procedure. Note that  existence of a periodic solution to Eq.~(\ref{EqQ1}) justifies the error estimate
(i.e. $O-$ term) in the asymptotic approximation (\ref{q=}).

Terms   of order $\omega$ in Eq.~(\ref{TxEq2Scl})-(\ref{PrdEq2Scl}), together with Eq. (\ref{q1=}), lead to equations
\begin{eqnarray}
&(\partial_\tau u_1-\nu_1\partial_{\xi\xi})u_1=2\nu_2\partial_{x\xi}u_0+\partial_x  (\kappa q_0+\sigma f)-\nu u_0-\partial_t u_0,\quad (\xi,\tau)\in \mathbb{T}^2;&
\label{EqU1}\\
 & \partial_\tau p_1+\partial_\xi(u_0p_1)-\nu_2\partial_{\xi\xi}p_1=2\nu_2\partial_{x\xi} p_0-\partial_t p_0-\partial_x(u_0p_0)-\partial_\xi(u_1p_0),\quad (\xi,\tau)\in \mathbb{T}^2.&
\label{EqP1}
\end{eqnarray}
On averaging Eqs.~(\ref{EqU1}) and (\ref{EqP1}) and using (\ref{p0=}), we obtain  the homogenized system (\ref{EqUSlw0})-(\ref{EqQSlw0}). Note that the resolving of the latter implies that Eqs.~(\ref{EqU1}) and (\ref{EqP1}) have periodic solutions $u_1$ and $p_1$. This, in turn, justifies $O-$terms in the approximations  (\ref{u=}) and (\ref{p=})).

Now we get back  to the above lemma. Before we start proving it in full generality, let us consider the particular case of $w=w(\xi)$. Then the
periodic solution to Eq.~(\ref{QEq}) has to  depend on a single variable $\xi$ and equation ~(\ref{QEq})
reduces to equation
$$
\partial_\xi(\eps\partial_\xi Q - wQ)=0.
$$
Its general solution takes the form
$$
Q=\exp\left(\eps^{-1}\int w(\xi)d\xi\right)\int\limits\exp\left(-\eps^{-1}\int w(\xi)d\xi\right)d\xi
$$
There is only one periodic solution among those given by this integral. To write it down explicitly, consider decomposition
$w(\xi)=\langle w(\xi)\rangle+\widetilde{w}(\xi)$,  and set
$$
E(\xi)=\exp((\eps\partial_\xi)^{-1}\widetilde{w}(\xi)).
$$
Then
$$
Q=AE(\xi)\int\limits_0^\infty\frac{ \mathrm{e}^{-s}d s}{E(\xi+\frac{s}{\mu})},\ \text{where}\  \mu=\frac{\langle w(\xi)\rangle}{\eps}\neq 0,\ A=\const
$$
and
$$
Q=AE(\xi),\ \langle w(\xi)\rangle=0,\ A=\const.
$$
Obviously, we can normalize the solution as required in (\ref{QAvr}) by  suitable choice of $A$ in both cases but, for our purposes, we need  to write down  the periodic solution in a form that covers  both cases. To do this, we consider the Fourier series
$$
E(\xi)=\sum\limits_{k} \psi_k\mathrm{e}^{ik\xi},\quad E^{-1}(\xi)=\sum\limits_{k} \phi_k\mathrm{e}^{ik\xi}.
$$
Then
$$
Q=AE(\xi)\left(\phi_0+\mu \sum\limits_{k\neq 0}\frac{\phi_k\mathrm{e}^{ik\xi}}{\mu-ik}\right),\quad
A^{-1}=\phi_0\psi_0+\mu \sum\limits_{k\neq 0}\frac{\phi_k\psi_k^*}{\mu-ik}
$$
where $\psi_k^*$ are the complex-conjugate Fourier coefficients.

Let us pass to the general case.
Let $\mathrm{H}$ be  space of the Fourier series in $\xi,\tau$ with  square-summable  coefficients
and let $\mathcal{L}:\mathrm{H}\to\mathrm{H}$ be operator defined by the left hand side of Eq.~(\ref{QEq}).
 We have to prove that
\begin{equation}\label{LmmExst}
    \dim\,\mathrm{Ker}\,\mathcal{L}=1, \ \langle w\rangle\neq 0\ \forall \ w\in \mathrm{Ker}\,\mathcal{L}\setminus\{0\}.
\end{equation}
Let $\mathcal{L}^*$ denote the operator adjoint  to $\mathcal{L}$. Define
$$
\breve{\mathcal{L}}^*=\mathcal{{J}}\mathcal{L}^*\mathcal{{J}}
$$
 where $\mathcal{J}:\mathrm{H}\to\mathrm{H}$ is the action of inversion $(\xi,\tau)\mapsto(-\xi,-\tau)$. Then
$$
\breve{\mathcal{L}}^*:\varphi\mapsto (\partial_\tau-\eps\partial_{\xi\xi})\varphi+w\partial_\xi\varphi.
$$
Notice that PDE
$$
(\partial_\tau-\eps\partial_{\xi\xi})\varphi+w\partial_\xi\varphi=0
$$
obeys the strong maximum and minimum principles (see e.g. \cite{Nrnbrg} or \cite{Lnds}). Hence,
$$
\mathrm{Ker}\,\breve{\mathcal{L}}^*=\{\varphi\equiv\const\}=\mathrm{Ker}\,\mathcal{L}^*.
$$
Applying unilateral strong maximum/minimum principles to PDE
$$
(\partial_\tau-\eps\partial_{\xi\xi})\varphi+w\partial_\xi\varphi=1
$$
shows that neither equation $\breve{\mathcal{L}}^*_v\breve{\psi}=1$ nor equation
$
{\mathcal{L}}^*_v\psi=1
$
 has a solution in $\mathrm{H}$.  Consequently,  the resolvent  operator $(\mathcal{L}^*-\lambda \mathcal{I})^{-1}$, $\lambda\in \mathbb{C}$  has a simple pole at  the origin. Moreover, since this resolvent is compact,  the pair of operators $\mathcal{L}^*$ and $\mathcal{L}$ obeys the Fredholm theorems. Hence $\dim\,\mathrm{Ker}\,\mathcal{L}=1$. Furthermore, if $\langle w\rangle=0$ for some $w\in \mathrm{Ker}\,\mathcal{L}\setminus\{0\}$, this would imply the existence of solution to equation $\mathcal{L}^*\psi=\const\neq 0$ but this contradicts to what has been proved above. This completes the proof.
\\
\textbf{Remark.} Let $\Pi^*$ denote spectral projector onto $\mathrm{Ker}\,\mathcal{L}^*$. Then the action of mapping (\ref{VActsOnR})-(\ref{EtaToP}) is identical to $\eta\mapsto \langle\Pi^*(\eta)\widetilde{u}_0\rangle$. Then the perturbation theory for linear operators implies that this mapping is differentiable and even analytic in a vicinity of origin, and that the derivative of this mapping denoted as  $\mathcal{V}^\prime(f)$ can be evaluated in accordance with (\ref{DffV})-(\ref{P1onXi})-(\ref{P0onXi}).
\end{document}